\documentclass[reprint,superscriptaddress, aps]{revtex4-1}

\usepackage{blindtext}
    \usepackage{amsmath}
    \usepackage{makeidx}
    \usepackage{amsfonts}
    \usepackage[ansinew]{inputenc}
    \usepackage[usenames,dvipsnames]{pstricks}
    \usepackage{subfigure}
    \usepackage{pst-grad} 
    \usepackage{pst-plot} 
    \usepackage[colorlinks,hyperindex]{hyperref}
    \hypersetup
    {
        colorlinks,%
        citecolor=black,%
        linkcolor=black,%
        urlcolor=black,%
    }

\usepackage{color}
\usepackage{dsfont}
\usepackage{graphicx}
\usepackage{subfigure}

\usepackage{dcolumn}
\usepackage{amsmath}    
\usepackage{verbatim}   
\usepackage{color}      
\usepackage{hyperref}   
\usepackage{amsfonts}
\usepackage{braket}
\usepackage{bm}
\usepackage{titlesec}
\usepackage{capt-of}
\usepackage{float}

\begin{document}


\title{Multi-dimensional photonic states from a quantum dot}

\author{J. P. Lee}\,
\affiliation{Toshiba Research Europe Limited, Cambridge Research Laboratory,\\
208 Science Park, Milton Road, Cambridge, CB4 0GZ, U.K.}
\affiliation{Engineering Department, University of Cambridge,\\
9 J. J. Thomson Avenue, Cambridge, CB3 0FA, U.K.}

\author{A. J. Bennett}
\affiliation{Toshiba Research Europe Limited, Cambridge Research Laboratory,\\
208 Science Park, Milton Road, Cambridge, CB4 0GZ, U.K.}
\email{anthony.bennett@crl.toshiba.co.uk}

\author{R. M. Stevenson}
\affiliation{Toshiba Research Europe Limited, Cambridge Research Laboratory,\\
208 Science Park, Milton Road, Cambridge, CB4 0GZ, U.K.}

\author{D. J. P. Ellis}
\affiliation{Toshiba Research Europe Limited, Cambridge Research Laboratory,\\
208 Science Park, Milton Road, Cambridge, CB4 0GZ, U.K.}

\author{I. Farrer}
\thanks{Current affiliation: Department of Electronic \& Electrical Engineering, University of Sheffield, Mappin Street, Sheffield, S1 3JD, U.K. }
\affiliation{Cavendish Laboratory, Cambridge University,\\
J. J. Thomson Avenue, Cambridge, CB3 0HE, U.K.}

\author{D. A. Ritchie}
\affiliation{Cavendish Laboratory, Cambridge University,\\
J. J. Thomson Avenue, Cambridge, CB3 0HE, U.K.}

\author{A. J. Shields}
\affiliation{Toshiba Research Europe Limited, Cambridge Research Laboratory,\\
208 Science Park, Milton Road, Cambridge, CB4 0GZ, U.K.}

\date{\today}%

\begin{abstract}
Quantum states superposed across multiple particles or degrees of freedom are of crucial importance for the development of quantum technologies. Creating these states deterministically and with high efficiency is an ongoing challenge. A promising approach is the repeated excitation of multi-level quantum emitters, which have been shown to naturally generate light with quantum statistics. Here we describe how to create one class of higher dimensional quantum state, a so called W-state, which is superposed across multiple time bins. We do this by repeated Raman scattering of photons from a charged quantum dot in a pillar microcavity. We show this method can be scaled to larger dimensions with no reduction in coherence or single photon character. We explain how to extend this work to enable the deterministic creation of arbitrary time-bin encoded qudits. 

\end{abstract}

\maketitle

\section{Introduction}

\begin{figure*}[t]
\includegraphics[width=15 cm]{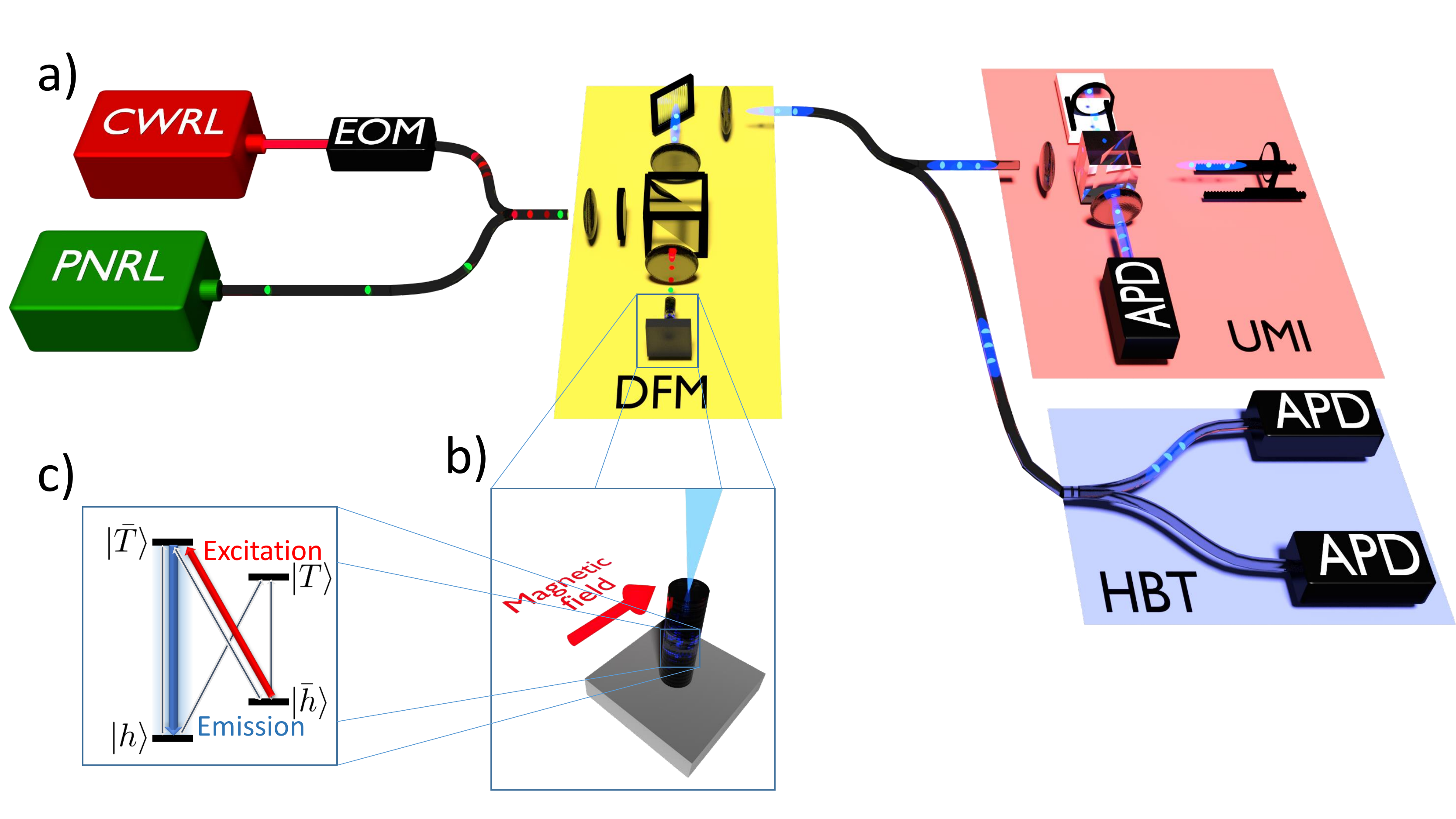}
\caption{ 	 
{\bf(a)} An illustration of the experimental setup. \textbf{(b)} The QD-micropillar system is cooled in a cryostat to 5K and placed in a Voigt geometry magnetic field. A pulsed non-resonant laser (PNRL) is used to apply short non resonant pulses and a continuous-wave resonant laser (CWRL) is modulated using an electro-optic modulator (EOM) to create a series of pulses.  The light is laser light is focused on the sample via a dark field microscope (DFM) that makes use of polarisation and spectral filtering to separate the output light emitted from the QD from the input laser light. The output light is then either directed into one of two detection setups. The output light can be directed into a Hanbury Brown and Twiss (HBT) setup consisting of a beam splitter and two avalanche photo-diodes (APDs). The output can also be directed into unbalanced Michelson interferometer (UMI).
{\bf(c)} The energy level diagram shows the states and allowed transitions of a single-hole charged quantum dot in a Voigt field. The highest energy vertical transition is cavity enhanced (shown as blue) and the $\ket{\bar{h}}\rightarrow\ket{\bar{T}}$ transition is driven by the CWRL (shown as red).}  
\label{Fig1}
\end{figure*}

Photonic states with multiple components are an essential resource for secure quantum relays \cite{collins2005quantum}, measurement based quantum computers \cite{raussendorf2003measurement} and quantum enhanced sensors \cite{giovannetti2004quantum}. Many early demonstrations of these technologies have made use of the polarisation degree of freedom, which is simple to produce and manipulate but is limited to two dimensions \cite{nagali2009quantum}. The challenge of extending these quantum states to greater dimensions in a scalable way, through adding extra quantum bits or degrees of freedom, promises new functionality and greater resistance to errors. In some cases replacing the `quantum bit' with a three dimensional qutrit or a d-dimensional qudit has clear advantages, for instance in quantum communication where the larger alphabet of characters allows transmission of more than one bit of classical information per photon \cite{PhysRevA.61.062308}. Such higher dimensional states can be encoded in the photon path, orbital angular momentum \cite{mair2001entanglement}, the radial degree of freedom \cite{karimi2014exploring}, temporal modes \cite{PhysRevLett.109.053602}, or perhaps most naturally in separate time bins \cite{PhysRevLett.82.2594, jayakumar2014time}. 

One interesting class of higher dimensional photonic state is the W-state, which is a state of the form:
\begin{equation}
\ket{W} = \frac{\ket{001} + \ket{010} + \ket{100}}{\sqrt{3}}.
\label{W-StateEqn}
\end{equation}

In the case of a single photon W-state, this can take the form of a single photon superposed across multiple modes. W-states have uses ranging from fundamental investigations of quantum mechanics to imaging and random number generation\cite{chaves2011feasibility, heaney2011extreme, gottesman2012longer, grafe2014chip}.

There are several approaches to realising higher dimensional photonic states. For example, it has been demonstrated that the widely studied parametric generation of entangled photon pairs can be extended to photon triplets \cite{hubel2010direct}. It is also possible to fuse together \cite{browne2005resource} or entangle \cite{o2003demonstration} smaller photonic states in a gate operation, but even for state-of-the-art technologies these approaches are probabilistic or have limited fidelities \cite{hacker2016photon}, which reduces the efficiency. An attractive option is to directly generate a complex photonic state from a single quantum emitter \cite{schon2005sequential, schon2007sequential}, which in principle allows the deterministic generation of multi-qubit photonic states. In fact, it has been argued that creating entangled photonic states on demand would be the final enabling technology in the development of a photonic quantum computer \cite{PhysRevLett.115.020502, rudolph2016optimistic}. Trapped spins in quantum dots have been shown to be suitable multi-level emitters for this purpose \cite{lindner2009proposal, Schwartzaah4758}. In addition, they have been used to demonstrate coherent spin manipulations \cite{press2008complete}, spin-photon entanglement \cite{de2012quantum, pmid:23151586} and distant entanglement between two spins \cite{delteil2015generationPub}. 

In this work, we use a cavity enhanced Raman transition in a quantum dot (QD) to show enhanced spin preparation and then to sequentially generate time-bin-encoded single photon W-states. Firstly, we demonstrate that a high Q-factor micropillar cavity allows us to observe cavity stimulated Raman emission. We use this effect to perform spin state preparation \cite{atature2006quantum} on a trapped hole spin over an order of magnitude faster than in the non cavity-enhanced case. We then demonstrate our scheme for W-state generation and show its scalability by producing photons superposed across up to four time bins. Finally, we explain how the techniques demonstrated in this work could allow the deterministic generation of arbitrary single photon time bin encoded states.  We anticipate that this capability will prove useful for single-mode quantum computation \cite{PhysRevLett.111.150501} and for maximising the key rates of QKD protocols over long distances \cite{1367-2630-17-2-022002, sasaki2014practical, takesue2015experimental}.

\section{Cavity-enhanced spin preparation}

We perform our experiments using a single-hole charged quantum dot in a micropillar cavity (FIG \ref{Fig1}b), a system which has been shown to be a good source of indistinguishable photons \cite{Bennette1501256}. Applying a Voigt geometry magnetic field results in a double-$\Lambda$ system. The magnetic field and temperature are tuned so that a single vertical transition is cavity-enhanced (FIG \ref{Fig1}c). The states of the system are aligned parallel/anti-parallel to the magnetic field. We label the hole spin states, $\ket{h}/\ket{\bar{h}} = \frac{\ket{\Uparrow}\pm\ket{\Downarrow}}{\sqrt{2}}$ and the trion states as $\ket{T}/\ket{\bar{T}} = \frac{\ket{\Uparrow\Downarrow\uparrow}\pm\ket{\Uparrow\Downarrow\uparrow}}{\sqrt{2}}$.

In order to maximise the number of operations that can be performed on a spin per unit time, it is desirable to perform spin state preparation as rapidly as possible. In the Voigt configuration the spin preparation typically takes several nanoseconds \cite{PhysRevLett.98.047401}.

\begin{figure}[h!]
\includegraphics[width=9 cm]{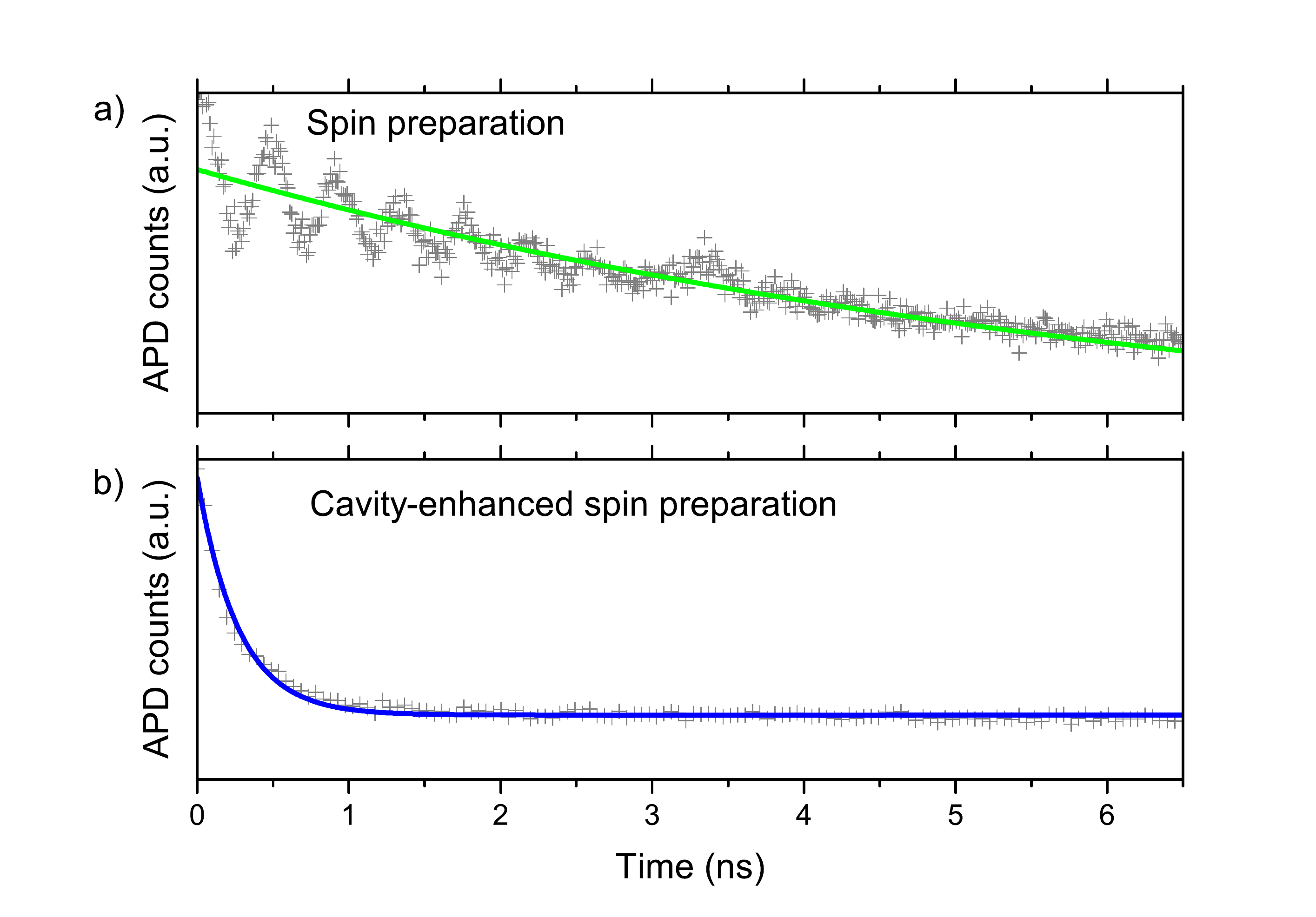}
\caption{{\bf a)} A graph showing the decrease in Avalanche Photo-diode (APD) counts over time due to spin pumping of trapped spin in a QD in a Voigt geometry magnetic field with no Purcell enhancement. The sinusoidal oscillations visible in the raw data are a result of Rabi oscillations between the levels driven by the resonant laser. The underlying exponential decay time (fitted green line) is $6.71 \pm 0.55$ ns.  {\bf b)} A graph showing the decrease in photon emission probability over time due to spin pumping in our Purcell enhanced system. The fitted exponential decay (blue line) has a decay time of $270 \pm 6$ ps. }
\label{Fig3}
\end{figure}

In our experiments we use a micropillar cavity to decrease the spin preparation time.  Driving the $\ket{\bar{h}} \rightarrow \ket{\bar{T}}$ transition we observe cavity-stimulated Raman emission \cite{sweeney2014cavity}. This increases the speed of spin preparation because system is more likely to decay via the enhanced vertical transition than the non-enhanced diagonal transition. As a result we can perform spin preparation in $270 \pm 6$ ps - over an order of magnitude faster than is typically seen for similar non-cavity enhanced systems (FIG \ref{Fig3}). We also note that the selective enhancement of one transition of the lambda system means that it can function as a cycling transition to be used for spin readout \cite{carter2013quantum}. The ability to perform rapid spin preparation and reliable spin readout in the Voigt geometry will help make trapped spins in quantum dots appealing candidates for stationary qubits and light matter interfaces.

The coherence time of the Raman scattered photons is determined by the coherence of the long lived ground state hole spin and the laser rather than the shorter coherence time of the exciton \cite{PhysRevLett.103.087406, PhysRevB.93.241302}. A further useful property of Raman emission is that the wavelength of the emitted photons can be tuned by tuning the wavelength of the excitation laser \cite{PhysRevLett.111.237403, sweeney2014cavity}. In our system we observe a tuning range of over 65 $\mu$eV. This would allow for the generation of indistinguishable photons from different sources - a key requirement for several quantum photonic technologies \cite{browne2005resource, knill2001scheme}.

\section{The sequential generation of time-bin encoded W-States}
Here we implement a scheme for the generation of a single photon W-state.
W-states are of particular interest as they represent one of two types of maximally entangled tripartite states and maintain their entanglement in the presence of dissipation \cite{dur2000three}. The concept can be generalized to include W-states with more than three qubits. 

\begin{figure}[h!]
\includegraphics[width=9 cm]{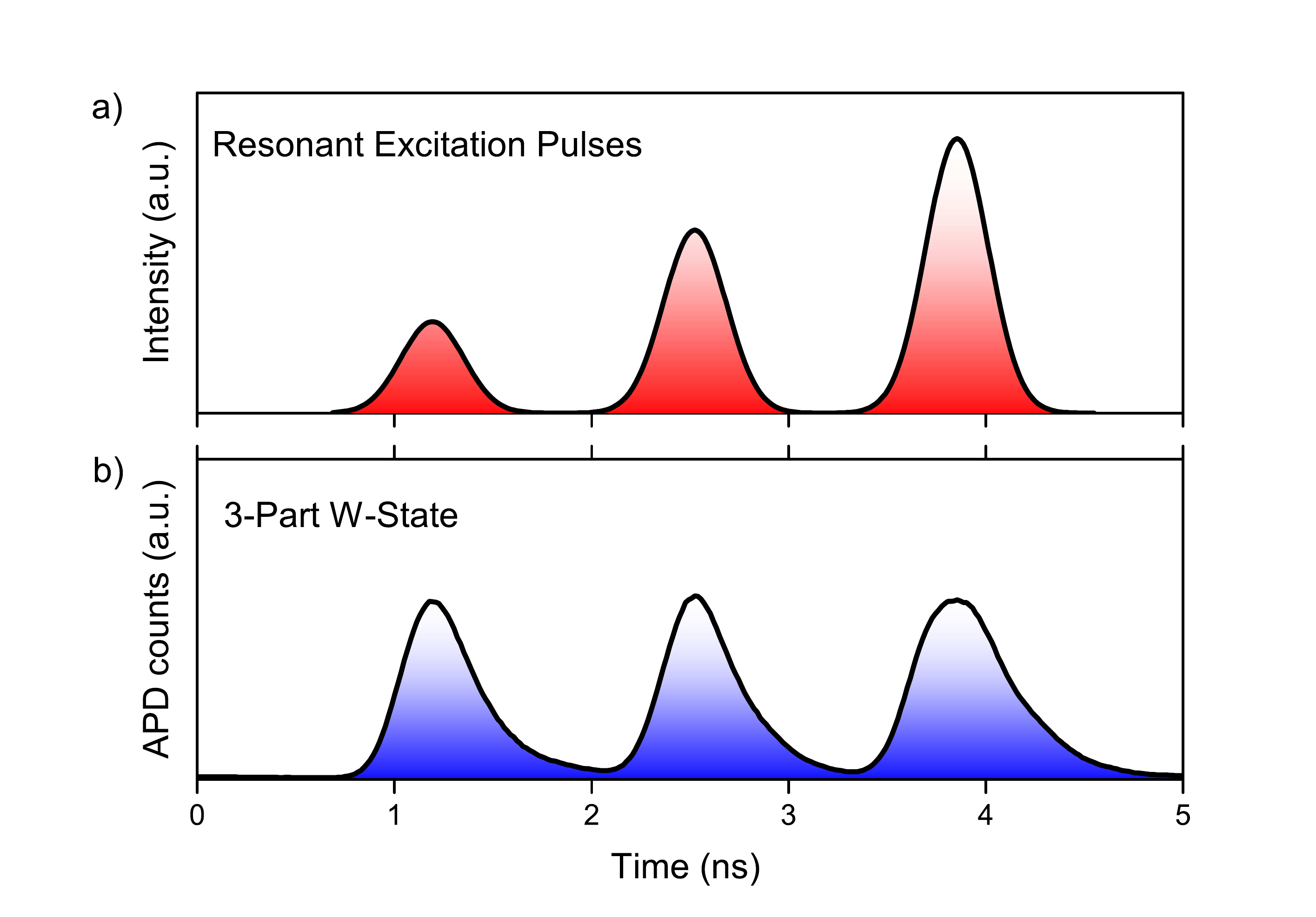}
\caption{{\bf a)} An illustration of the pulse sequence used to create a 3 qubit single photon W-State. The series of three pulses is used to drive the Raman transition and create W-state photon. {\bf b)} A time resolved measurement of the collected light. Spectral filtering ensures that the collected light is at the wavelength of the cavity enhanced transition.}  
\label{Fig4}
\end{figure}

We use a series of weak resonant pulses (shown in FIG \ref{Fig4}a) to drive the $\ket{\bar{h}} \rightarrow \ket{\bar{T}}$ transition (shown as the red arrow in FIG \ref{Fig1}c). The power of these pulses is tuned so that they each have the same probability of driving the transition and generating a photon. When the $\ket{\bar{h}} \rightarrow \ket{\bar{T}}$ transition is driven, the system preferentially decays vertically (shown as the blue transition in FIG \ref{Fig1}c). The emitted photon has an equal probability of being measured in each time bin (FIG \ref{Fig4}b). The light produced by this mechanism is emitted as single quanta because the detection of a photon at the wavelength of the enhanced transition heralds the preparation of the hole spin in the $\ket{h}$ state. Once in the $\ket{h}$ state the system cannot be re-excited by the laser driving the $\ket{\bar{h}} \rightarrow \ket{\bar{T}}$ transition until a spin-flip occurs. Spin-flips typically take three orders of magnitude longer than the W-state generation process and so are not of concern here \cite{heiss2007observation}. In contrast, the single photon emission that is usually observed from quantum dots relies on the spontaneous decay time of a transition being greater than the applied pulse length \cite{:/content/aip/journal/apl/82/14/10.1063/1.1563050}.

The setup for our scheme is shown in FIG \ref{Fig1}a. A non-resonant pulse ensures that there is a non-zero population in the $\ket{\bar{h}}$ state. Then resonant pulses produced using a continuous wave laser and an electro-optic modulator (EOM) are used to drive the Raman transition.

\begin{figure*}[t!]
\begin{minipage}[h]{0.45\textwidth}
\includegraphics[width=8 cm]{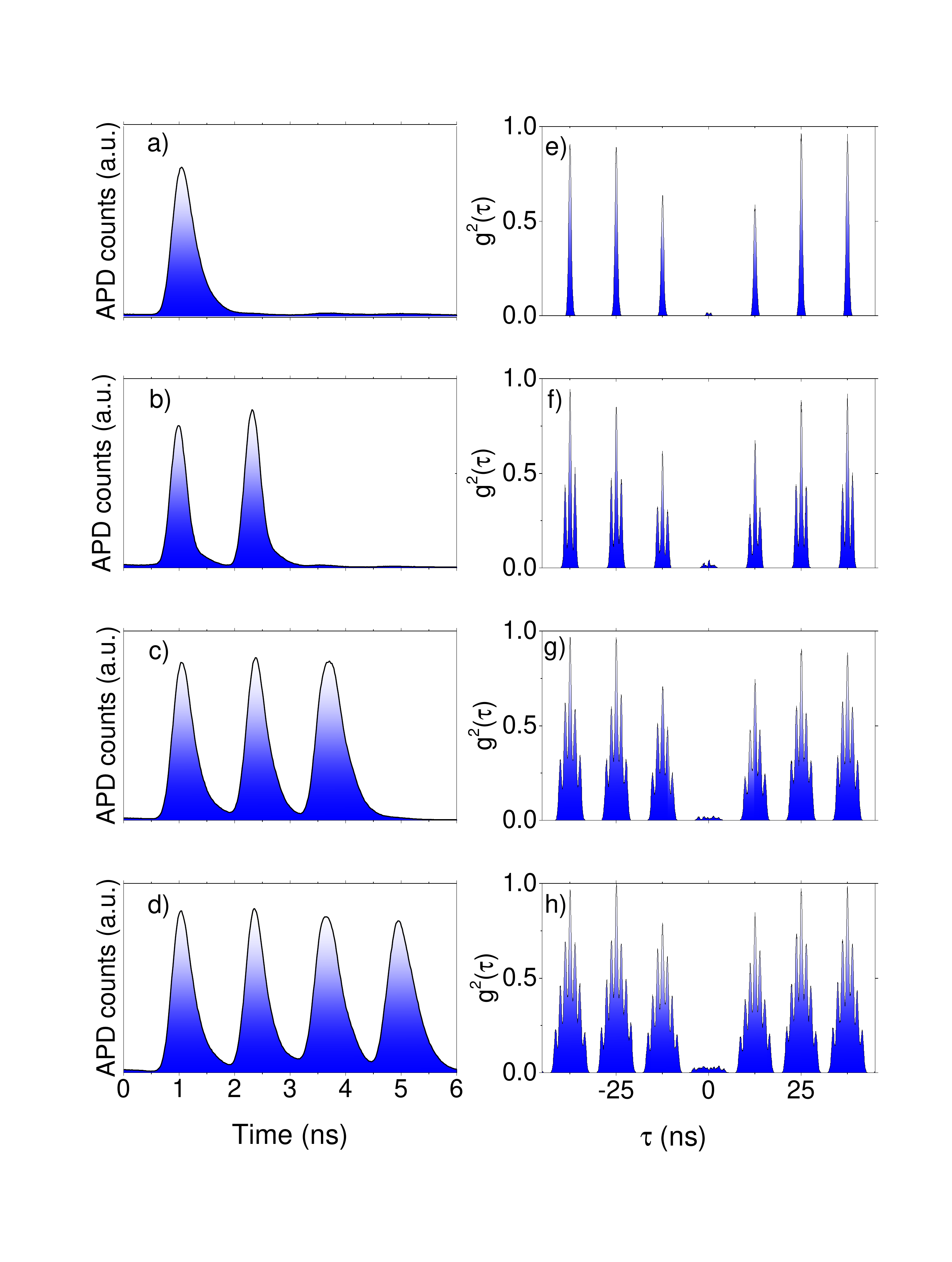}
\end{minipage}
\begin{minipage}[h]{0.45\textwidth}
\includegraphics[width=8 cm]{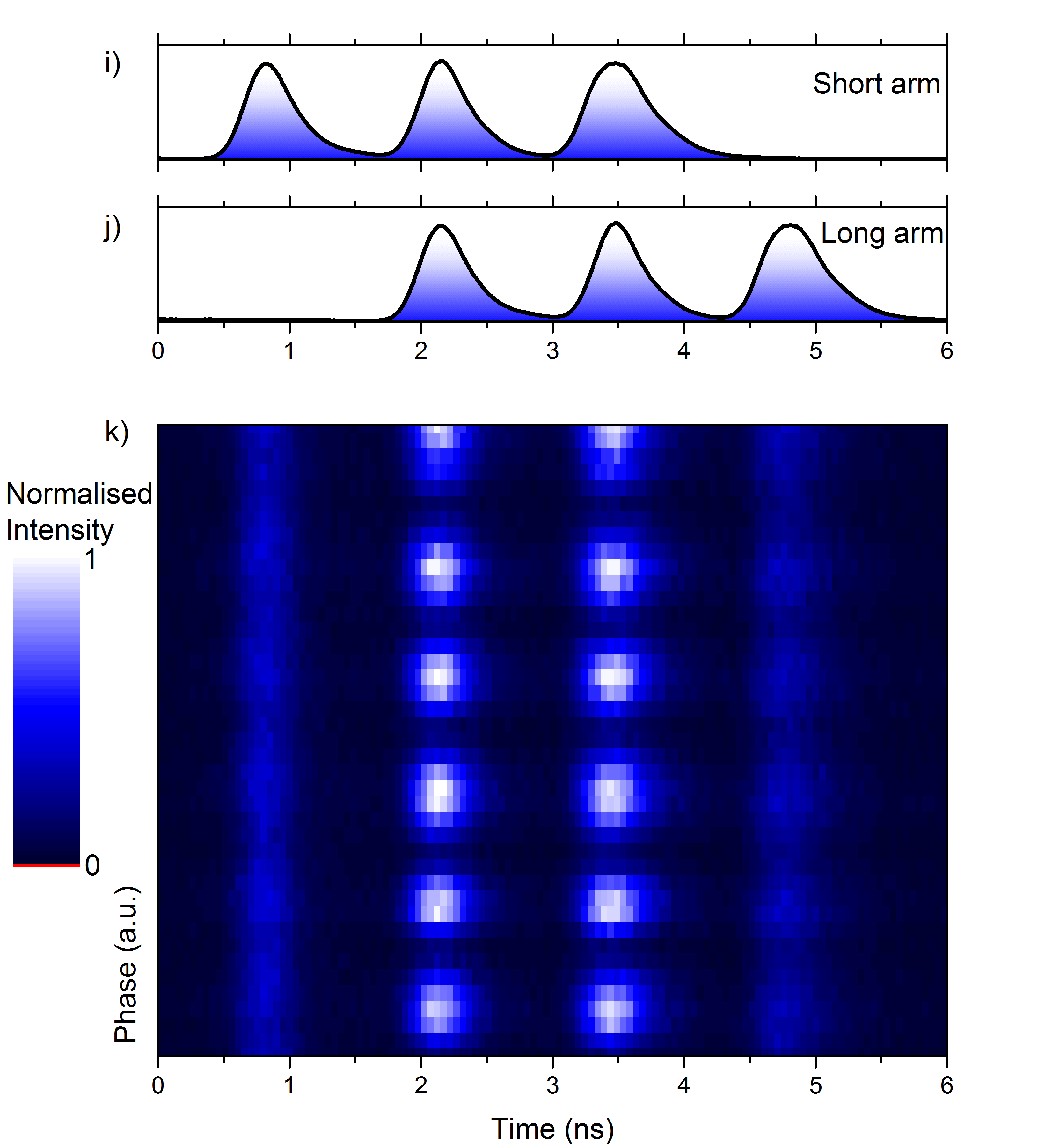}

\end{minipage}
\caption{{\bf a-d)} Time resolved measurements of the produced photons superposed across 1, 2, 3 and 4 time-bins. {\bf e-h)} Second-order correlation function measurements on the produced W-state photons confirm their single-photon nature. {\bf i, j)} Time resolved measurements showing the single-time-bin offset introduced by the difference in the short and long paths of the unbalanced Michelson interferometer. {\bf k)} A time resolved interference measurement showing interference between neighbouring time-bins of a single photon superposed across three time-bins.}
\label{Fig5}
\end{figure*}

First, we use a single resonant pulse and second order correlation function measurement to demonstrate that the system can act as a single photon source. FIG \ref{Fig5}a shows a time resolved measurement of the output photons and FIG \ref{Fig5}e shows the second order correlation function measurement. The low $g^{(2)}(0) \approx 0.02$ indicates that the output light is primarily composed of single photons. 

We then use a two pulse sequence to create a photon superposed across two time bins (FIG \ref{Fig5}b). Extending this to a three pulse sequence we generate a single photon three-part W-state (FIG \ref{Fig5}c). Finally, we use a four pulse sequence to create a four part W-state and demonstrate the scalability of our scheme (FIG \ref{Fig5}d). 
The output light for all of the pulse sequences have a greatly reduced peak in the second order correlation function at $\tau=0$, indicating that the output is dominated by single photon emission (FIG \ref{Fig5}e-h). 

We also observe anti-bunching over tens of nanoseconds; the peaks at $\tau = \pm 12.5$ ns are lower than the other peaks away from $\tau = 0$. This indicates that the non-resonant pulse does not completely randomise the hole spin state meaning that a photon is less likely to be generated if a photon has been produced during the preceding excitation sequence. This effect could be removed by deterministic preparation of the spin in the $\ket{\bar{h}}$ state prior to the incidence of the resonant laser pulses.


\begin{table}[ht]
\caption{Coherent single-photon states} 
\centering 
\begin{tabular}{c c c c} 
\hline\hline 
\# Time-bins & $g^{(2)}(0)$ & Visibility \\ [0.5ex] 
\hline 
1 & $0.0233 \pm 0.0009$ & -  \\ 
2 & $0.0463 \pm 0.0012$ & $65.6 \pm 4.3\%$  \\
3 & $0.0309 \pm 0.0004$ & $69.6 \pm 5.4\%$  \\
4 & $0.0577 \pm 0.0004$ & $67.7 \pm 5.7\%$  \\ [1ex] 
\hline 
\end{tabular}
\label{table} 
\end{table}

Finally, we probe the coherence between neighbouring time-bins of the three part W-state by time resolving the output of an unbalanced Michelson interferometer. This measurement shows that the photons are in a coherent superposition between different time-bins and not simply emitted into one time-bin or another probabilistically. The long arm of the interferometer delays the light by one time-bin relative to the short arm (FIG \ref{Fig5}i\&j). This means that we expect to see interference between time-bin 1 and time-bin 2, and between time-bin 2 and time-bin 3 of a single 3-time-bin photon. By varying the phase difference between the two arms we observe interference between the overlapping time-bins (FIG \ref{Fig5}k). The interference measurements were also performed for the 2 and 4 part W-states; the visibilities obtained by fitting a sinusoidal function to the measured intensities of the overlapping time bins as a function of phase are shown in Table \ref{table}. We attribute the deviation from perfect interference visibility to the decoherence of the hole spin. Using the 3-part W-state interference visibility measurement we estimate the hole $T_{2}^{*}$ time to be $\sim 3.7$ ns - within the range of previously measured hole inhomogeneous dephasing times \cite{PhysRevB.89.075316}.

\section{Proposal for the deterministic generation of single photon W-States}

As quantum dots in high Q cavities can act as deterministic photon sources \cite{nowak2014deterministic}, we propose an extension to our scheme to generate generate W-States deterministically. Our proposal is as follows:

\begin{enumerate}
\item Ensure that the system in the state $\ket{\bar{h}}$ by first preparing the $\ket{h}$ state as demonstrated earlier, then using an off-resonant pulse to perform a $\pi$-rotation of the hole spin\cite{press2008complete}.
\item Apply a $\frac{\pi}{3}$ pulse to the diagonal transition indicated in FIG \ref{Fig3}a. The resulting Raman emission leaves us with the state: $\sqrt{\frac{2}{3}}\ket{\bar{h}}\ket{0_{\tau=1}} + \frac{1}{\sqrt{3}}\ket{h}\ket{1_{\tau=1}}$, where $\ket{0_{\tau=1}}$ $(\ket{1_{\tau=1}})$ indicates the absence (presence) of a photon in time-bin 1.
\item A $\frac{\pi}{2}$ pulse leaves us with the state: $\frac{1}{\sqrt{3}}\ket{\bar{h}}\ket{0_{\tau=2} 0_{\tau=1}} + \frac{1}{\sqrt{3}}\ket{h}\ket{1_{\tau=2} 0_{\tau=1}} + \frac{1}{\sqrt{3}}\ket{h}\ket{0_{\tau=2} 1_{\tau=1}}$.
\item A final $\pi$ pulse leaves us with the state $\frac{1}{\sqrt{3}}\ket{h}\ket{1_{\tau=3} 0_{\tau=2} 0_{\tau=1}} + \frac{1}{\sqrt{3}}\ket{h}\ket{0_{\tau=3} 1_{\tau=2} 0_{\tau=1}} + \frac{1}{\sqrt{3}}\ket{h}\ket{0_{\tau=3} 0_{\tau=2} 1_{\tau=1}}$.
\end{enumerate}

Ignoring the hole state and considering the state of the photon in isolation it is clear that this gives us with a photon in the state $\frac{1}{\sqrt{3}}(\ket{001} + \ket{010} + \ket{100})$ - a single photon W-state.

\section*{Conclusions and outlook}
We have generated time-bin encoded single photon W-states from the cavity enhanced Raman emission of a quantum dot. The use of an EOM in this scheme allows us to make use of flexible electronic triggering \cite{Dada:16} to determine the probability amplitudes of the photon for each time bin.


A phase modulator could be used to control the relative phase of the laser between two resonant pulses to create a qubit suitable for time-bin-encoded quantum key distribution \cite{1367-2630-15-5-053007}. This phase modulation could also be achieved by using a detuned pulse to rotate the hole spin about the z-axis of the Bloch sphere \cite{delteil2015generationPub}. In combination with the demonstrated ability to control the amplitude of each time bin, this would allow the creation of arbitrary single photon time-bin-encoded states, which have uses in quantum communication and computation. 

Using continuously varied excitation rather than pulsed excitation would allow the generation of arbitrarily shaped photons \cite{1367-2630-13-10-103036}. This, in addition to phase modulation would enable to encode quantum information in the temporal mode of a photon \cite{PhysRevX.5.041017}. In combination with the wavelength tuning made possible by the cavity enhanced Raman emission process this will allow the generation of photons from QD sources that are indistinguishable from photons from other light sources such as lasers and trapped atoms. We expect our results to pave the way for solid state sources of on-demand photonic qubits and efficient interfaces between quantum dots and other quantum optical systems.

\section*{Methods}
The experiments were performed using an InAs quantum dot in a GaAs/AlGaAs micropillar cavity with a Q-factor of $\sim 7500$. The dot-cavity system was cooled to 5K. A pulsed non-resonant 850 nm laser is used to non-resonantly excite the quantum dot and ensure that there is a finite population in the $\ket{\bar{h}}$ state before the any resonant pulses are applied. The resonant pulses are generated using a continuous wave diode laser at $\sim 940$ nm and electro-optic modulator. The resonant pulses are separated from the emitted light by both polarisation and spectral filtering. The emitted light is the detected by fibre coupled avalanche photodiodes. To record the HBT measurements we time-tag each photon, this allows us to remove the effect of photons generated from the non-resonant pulses on the correlation measurements through temporal filtering. The unbalanced Michelson interferometer used was kept at a constant temperature in order to minimise drift.

\section*{Acknowledgments}
The authors acknowledge funding from the EPSRC for MBE system used in the growth of the micropillar cavity.
J. L. gratefully acknowledges financial support from the EPSRC CDT in Photonic Systems Development and Toshiba Research Europe Ltd.

\section*{Data Access}
The experimental data used to produce the figures in this paper is publicly available at [Information will be made available on Cambridge Data Repository before publication].


%

\end{document}